\documentclass[runningheads]{llncs}
\usepackage{capt-of}% or \usepackage{caption}
\usepackage{booktabs}
\usepackage{varwidth}

\usepackage{enumitem}
\usepackage{acro}
\usepackage[T1]{fontenc}
\usepackage{graphicx}
\usepackage{booktabs}
\usepackage{siunitx}
\usepackage{amsmath}
\usepackage{amssymb}
\usepackage{url}
\usepackage{hyperref}
\hypersetup{
    colorlinks=true,
    linkcolor=blue,
    filecolor=magenta,    
}
\usepackage{pythonhighlight}

\DeclareAcronym{mse}{
 short=MSE,
 long=mean squared error,
}

\DeclareAcronym{mri}{
 short=MRI,
 long=magnetic resonance imaging,
}

\DeclareAcronym{ct}{
 short=CT,
 long=computed tomography,
}

\DeclareAcronym{ae}{
 short=AE,
 long=autoencoder,
}

\DeclareAcronym{dae}{
 short=DAE,
 long=denoising autoencoder,
}

\DeclareAcronym{mae}{
 short=MAE,
 long=masked autoencoder,
}

\DeclareAcronym{vit}{
 short=ViT,
 long=visual transformer,
}

\DeclareAcronym{maemi}{
 short=MAEMI,
 long=masked autoencoder for medical imaging,
}

\DeclareAcronym{ad}{
 short=AD,
 long=anomaly detection,
}

\DeclareAcronym{gan}{
 short=GAN,
 long=generative adversarial network,
}

\DeclareAcronym{oct}{
 short=OCT,
 long=optical coherence tomography,
}

\DeclareAcronym{dce}{
 short=DCE,
 long=dynamic contrast enhanced,
}

\DeclareAcronym{auroc}{
 short=AUROC,
 long=area under the receiver operating characteristics curve,
}

\DeclareAcronym{ap}{
 short=AP,
 long=average precision,
}

\DeclareAcronym{sota}{
 short=SOTA,
 long=state of the art,
}

\DeclareAcronym{fpr}{
 short=FPR,
 long=false positive rate,
}

\DeclareAcronym{tn}{
 short=TN,
 long=true negative,
}

\DeclareAcronym{tp}{
 short=TP,
 long=true positive,
}

\DeclareAcronym{nfs}{
 short=NFS,
 long=non-fat saturated,
}

\DeclareAcronym{fs}{
 short=FS,
 long=fat saturated,
}

\begin{document}
\title{3D Masked Autoencoders with Application to Anomaly Detection in Non-Contrast Enhanced Breast MRI}
\titlerunning{MAEMI for Anomaly Detection in Breast MRI}
\authorrunning{Lang et al.}
\author{
  Daniel M. Lang\inst{1,2} \and
  Eli Schwartz\inst{3,4} \and
  Cosmin I. Bercea\inst{1,2} \and
  Raja Giryes\inst{4} \and
  Julia A. Schnabel\inst{1,2,5}
\institute{
  Helmholtz Munich, Germany \and
  Technical University of Munich, Germany \and
  IBM Research AI, Israel \and
  Tel-Aviv University, Israel \and
  King's College London, United Kingdom
}
\\
\email{daniel.lang@tum.de}
}

\maketitle

% Abstract
% -----------------------------------------------------------------------------------------
\begin{abstract}
Self-supervised models allow (pre-)training on unlabeled data
and therefore have the potential to overcome the need for large annotated
cohorts.
One leading self-supervised model is the \ac{mae} which was developed on
natural imaging data. The \ac{mae} is
masking out a high fraction of \ac{vit} input patches, to then
recover the uncorrupted images as a pretraining task.
In this work, we extend MAE to perform anomaly detection on breast \ac{mri}.
This new model, coined \ac{maemi} is
trained on two non-contrast enhanced \ac{mri} sequences, aiming at lesion
detection without the need for intravenous injection of contrast media and
temporal image acquisition.
During training, only non-cancerous images are presented to the model,
with the purpose of localizing anomalous tumor regions during test time.
We use a public dataset for model development.
Performance of the architecture is evaluated in reference to subtraction images
created from \ac{dce}-\ac{mri}.

\end{abstract}
% -----------------------------------------------------------------------------------------
% Introdution
% -----------------------------------------------------------------------------------------
\section{Introduction}
Annotation of medical data requires expert knowledge
or labor-intensive testing methods, leading to high curation costs.
Therefore, labeled medical imaging datasets are typically several orders of magnitude
smaller than datasets generally encountered in computer vision.
Deep learning networks require large amounts of data to be trained,
making deployment of models in the medical domain cumbersome \cite{ching2018opportunities}.
Self-supervised learning aims at model development in the absence of labeled
examples and has the power to overcome those limiting factors \cite{sun2017revisiting}. 
A pretraining task is utilized to induce prior knowledge into the model,
which will then be fine-tuned for the respective downstream task of interest.
One leading self-supervised approach is the \acf{mae} \cite{he2022masked}, which was developed
on natural imaging data.
\Ac{mae} is a transformer based \ac{ae} model that randomly removes a high fraction
of its input patches, with the intention to recover the uncorrupted
images as a self-supervised task.

In addition to constraints in data acquisition, medical datasets are also often highly imbalanced,
featuring a skewed proportion of \textit{healthy} and \textit{unhealthy} examples.
\Ac{ad} models are designed to identify rare, uncommon elements that differ significantly
from \textit{normal} cases.
In the medical domain, such models are employed to distinguish \textit{abnormal} patterns
of unhealthy examples from normal patterns of healthy cases.

Self-supervised anomaly detection combines both training strategies,
aiming to identify abnormal cases without the requirement for labeled examples.
This can be achieved by reconstruction-based methods
\cite{bergmann2018improving,gong2019memorizing}.
Models are trained to recover their input, while restrictions on the
architecture are applied.
Such restrictions can be imposed by information bottlenecks \cite{chow2020anomaly} or the alteration
of input images by application of noise \cite{kascenas2022denoising,Wyatt_2022_CVPR}
or removal of image parts \cite{zavrtanik2021reconstruction,yan2021learning}.
During training, normal examples are shown to the model.
In this way, the model is only able to reconstruct
image parts stemming from the normal distribution reasonably well,
while abnormal image parts result in higher error rates
that can be utilized to generate anomaly maps during test time.
\textit{Schwarz et al.} \cite{schwartz2022maeday} modified \ac{mae}
to perform anomaly detection on natural imaging data.
Most \ac{ad} models in the area of medical imaging have been
developed on \ac{mri} of the brain, e.g.
\cite{baur2021autoencoders,bercea2022federated,kascenas2022denoising}.
Further areas of application include,
e.g. chest X-ray, \ac{oct} and mammography \cite{tschuchnig2022anomaly}.

We aim at model development on breast \ac{mri},
which is the most sensitive breast cancer imaging method
\cite{leithner2019abbreviated}, applied for tumor
staging but also cancer screening.
\Ac{dce}-\ac{mri} refers to the acquisition of images before,
during and after intravenous injection of contrast media,
which improves the signal intensity of neoangiogenically induced vascular
changes that allows for better detection of lesions \cite{turnbull2009dynamic}.
However, long scan times and high costs limit widespread use of the
technique, leading different studies to investigate the ability to
abbreviate \textit{contrast enhanced} breast \ac{mri} protocols
\cite{leithner2019abbreviated}.
We demonstrate the capability of self-supervised models for
anomaly detection on \textit{non-contrast enhanced} breast \ac{mri},
which reduces the number of required image sequences dramatically and
therefore results in even faster image acquisition.
Moreover, no intravenous injection of contrast media is needed,
which is known to be able to cause side effects \cite{hasebroock2009toxicity}.

\subsubsection{Contribution}
In this work we remodel \ac{mae} and extended and further develop the
approach of \textit{Schwarz et al.} \cite{schwartz2022maeday},
enabling self-supervised anomaly
detection on 3D multi-spectral medical imaging data.
To do so, we advance the definition of input patches and positional embedding
of the \ac{vit} architecture and refine the random masking strategy of
\textit{He et al.} \cite{he2022masked}.
We then train the model on non-contrast enhanced breast \ac{mri}.
During training only healthy, non-cancerous breast \acp{mri} are shown to the model,
aiming to identify breast lesions as anomalies during test time.
To the best of our knowledge, we are the first to make the following contributions:
\begin{itemize}
    \item We extend and further refine \acp{mae} to perform
    anomaly detection on 3D multi-spectral medical imaging.
    \item We investigate the capability of self-supervised anomaly 
    detection to identify pathologies in breast \ac{mri}.
    \item We assess the performance of anomaly detection algorithms in reference
    to \ac{dce}-\ac{mri} subtraction images.
    Paving the way for a more widespread use of \ac{mri} in breast
    cancer diagnosis.
\end{itemize}
% -----------------------------------------------------------------------------------------
%  Related Work
% -----------------------------------------------------------------------------------------
\section{Related Work}
The ability of deep convolutional \acp{ae} to perform reconstruction based
self-supervised anomaly detection on
imaging data has been investigated by several studies, see e.g.
\cite{baur2021autoencoders,bercea2022federated,somepalli2021unsupervised}.
\textit{Kascenas et al.} \cite{kascenas2022denoising} trained a denoising
\ac{ae} on brain \ac{mri}, such that unhealthy pathologies were removed
during test time.
\textit{Zavrtanik et al.} \cite{zavrtanik2021reconstruction} developed a
convolutional \ac{ae}, masking part of the input to
perform inpainting on natural imaging and video data.

\Ac{mae} has been developed on 2D natural images,
to be finetuned on a classification problem.
In the context of classification the approach has been modified in several different ways.
\textit{Feichtenhofer et al.} \cite{feichtenhofer2022masked} enhanced the model to do
classification on natural video data.

\textit{Prabhakar et al.} \cite{prabhakar2023vit}
improved the initial \ac{mae} architecture,
by incorporation of a contrastive and a auxiliary loss term,
to perform classification on brain \ac{mri}.
Our model relies on image reconstruction and computation
of a voxel-wise difference in the downstream task.
Therefore, we use unaltered \ac{mse} as a loss.

Due to its high masking ratio, anomalies are likely to be removed by \ac{mae}.
Which led \textit{Tian et al.} \cite{tian2022unsupervised} to employ the model
on anomaly detection in 2D colonoscopy and X-ray data.
They introduced \textit{memory-augmented self-attention} and a
\textit{multi-level cross-attention operator}
in the underlying \ac{vit} architecture, to limit dependency on random masking.
In contrast, we train our model on multi-spectral 3D data following the
principle strategy of \textit{Schwartz et al.} \cite{schwartz2022maeday}.
The approach does not rely on any modifications in the \ac{vit} architecture,
and maximizes the likelihood for the anomaly to be removed by application
of a high number of random masks.
We extend and further develop the model trained on natural
imaging data to be able to
handle multi-spectral volumetric medical imaging data.

\Ac{mae} based \acl{ad} models employ a problem independent pretraining task,
recovering pseudo-normal images from the masked input.
\Acp{dae} used for \ac{ad} are aiming to achieve the same effect by
pretraining on noise removal.
However, selection of noise has to fit the distribution of possible anomalies
for the approach to succeed. 
Hence, a model trained on one specific task is very unlikely to
succeed reasonably well in another problem setting.
In contrast to that, the ability of \ac{mae} based anomaly detection
to succeed in modified settings has been proven by \cite{schwartz2022maeday},
achieving \ac{sota} performance on few- and zero-shot problems.

Identification of lesions in breast \ac{mri} has only been performed
by \textit{supervised} models so far.
\textit{Maicas et al.} \cite{maicas2017deep} trained a deep Q-network for
breast lesion detection, \textit{Ayatollahi et al.} modified RetinaNet and
\textit{Herent et al.} \cite{herent2019detection}
utilized a 2D ResNet50. 
Notably, all of those approaches were trained on \ac{dce}-\ac{mri} data,
relying on injection of contrast media.
Whereas, we perform self-supervised anomaly detection
on non-contrast enhanced \ac{mri}.
% -----------------------------------------------------------------------------------------
%  Dataset
% -----------------------------------------------------------------------------------------
\section{Dataset}
We use the public Duke-Breast-Cancer-MRI cohort \cite{saha2021dynamic,saha2018machine}
from The Cancer Imaging Archive \cite{clark2013cancer}.
The set includes axial breast \ac{mri} data of 922 patients comprising a non-fat saturated
pre-contrast T1-weighted sequence, a fat-saturated T1-weighted pre-contrast sequence
and several post-contrast fat-saturated T1-weighted sequences.
Tumor lesion annotations were given in the form of bounding boxes.
Furthermore, a U-Net model for generation of breast tissue masks was provided \cite{UNET}.
All images were standardized to the same voxel spacing of $0.75 mm \times 0.75 mm \times 1.0 mm$,
cropped to involve the chest area only, and normalized to a mean and standard deviation
value of $0.5$ and $0.25$ per image.
Cases involving bilateral breast cancer were removed from the cohort.
Thus, each of the remaining patients exhibited one breast that contained a tumor lesion,
treated as \textit{abnormal/unhealthy}, and another breast not affected
by cancer, treated as \textit{normal/healthy}.
The dataset was split into a training set of 745 patients, a validation set
of 50 patients and a test set of 100 patients.
Notably, no \textit{normal - abnormal} pairs of the same patient were involved in
different datasets.
Due to the high memory requirements of transformer models, MRI-patches
of size $240 \times 168 \times 8$ voxels in lateral-posterior-superior (LPS)
directions were cropped from both \ac{mri} sequences to be then divided into
ViT-patches and processed by the encoder.

Following clinical routine, subtraction images $S$ between the image acquired before $I_{\text{pre}}$ and
all of the $m$ images acquired after injection of contrast media $I_{\text{post}}^k$ were computed:
\begin{equation}
  S = \frac{1}{m} \sum_{k=0}^m\left(I_{\text{pre}} - I_{\text{post}}^k\right)^2 \ast \text{min}_{3\times3\times2},
\end{equation}
with a minimum filter of size $3\times3\times2$ applied for noise removal.
% -----------------------------------------------------------------------------------------
% Method
% -----------------------------------------------------------------------------------------
\section{Method}
\label{sec:methods}
\begin{figure}
  \centering
  \includegraphics[width=0.9\textwidth]{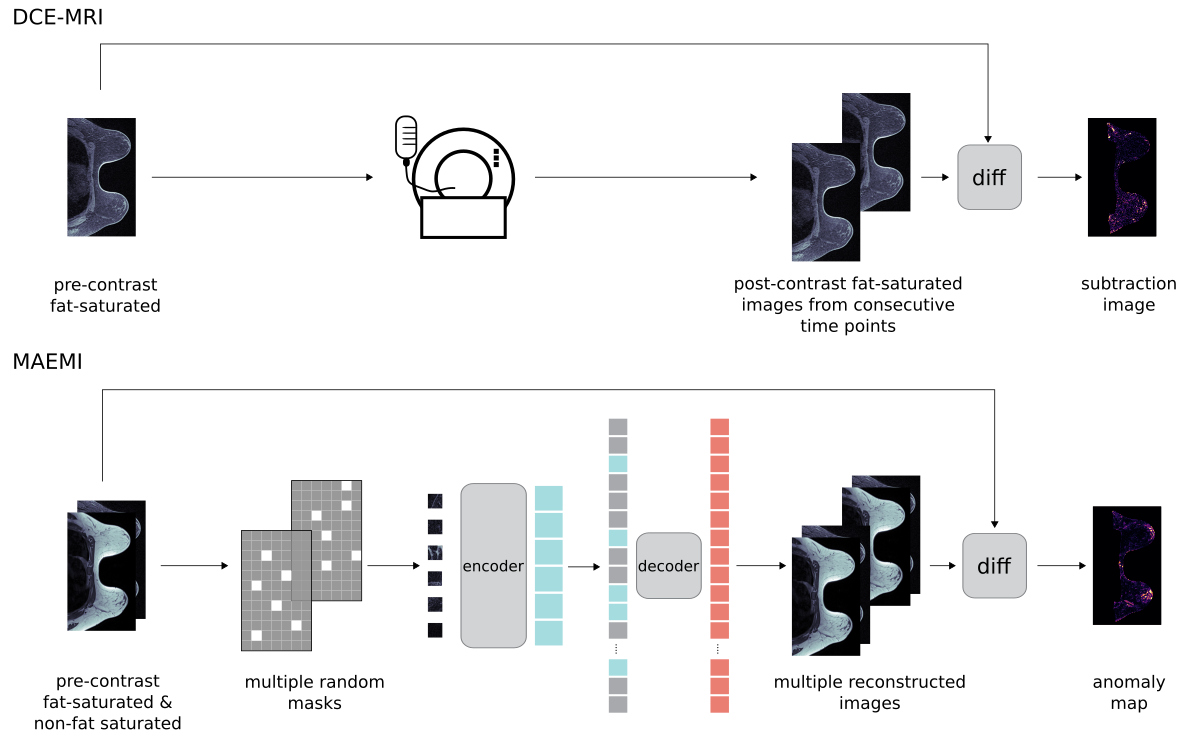}
  \caption{
    \ac{dce}-\ac{mri} imaging vs. \ac{maemi}.
    For \ac{dce}-\ac{mri}, several \acp{mri} before, during and
    after injection of contrast media are acquired.
    \Ac{maemi} uses different random masks for generation
    of pseudo-healthy recovered images.
    Both methods construct error maps  by calculation of the \acl{mse}
    difference between each of the post-contrast/reconstructed images
    with the pre-contrast/ uncorrupted image.
  }
  \label{fig:model}
\end{figure}

A scheme of our model can be seen in Figure \ref{fig:model}.
We modified the \ac{vit} architecture of \cite{dosovitskiy2020image}
to 3D multispectral MRI,
i.e. the positional embedding and definition of ViT-patches was redefined and
enhanced to incorporate a third dimension.
This 3D-\ac{vit} was then embedded in the \ac{mae} approach of \cite{he2022masked}.
To do so, generation of mask tokens and random masking of ViT-patches had to be
remodeled.
The approach coined \acf{maemi} uses a 3D-\ac{vit}
with 12 transformer blocks and a embedding dimension of 768 as encoder
while for the decoder a embedding dimension of 384 and a depth of 4 has been chosen.

In addition, we further improved the anomaly detection model of
\cite{schwartz2022maeday}.
\Acp{mri} were processed by patches, such that  memory requirements of
the transformer based architecture could be reduced, enabling exploitation of
whole MRI volumes.
A overlapping patch scheme has been chosen, in order to reduce
artifacts on patch borders.
Anomaly maps were generated from two input sequences, i.e. \ac{nfs} and \ac{fs}.

Error maps per input sequence $E^{\text{seq}}_i$ were constructed by
the \ac{mse} between the reconstructed patches $R^{\text{seq}}_i$ and the
unmasked \ac{mri}-patch $I^{\text{seq}}$:
\begin{equation}
    E_i^{\text{seq}} = (I^{\text{seq}} - R^{\text{seq}}_i)^2 \ast \text{min}_{3\times3\times2},
\end{equation}
with a minimum filter of size $3\times3\times2$ applied for further reduction of artifacts.
On a voxel level, final error scores were computed by the mean
value of all patch predictions.
Error maps of both multi-spectral input sequences, \ac{nfs} and \ac{fs},
were then summed up and convolved with the same minimum filter as before:
\begin{equation}
    E = \frac{1}{2} \left( E^{\text{NFS}} + E^{\text{FS}}\right)
    \ast \text{min}_{3\times3\times2},
\end{equation}
for generation of a final MR image level anomaly map.

\paragraph{Training Specifics}
During training, only MRI-patches of healthy breasts, containing
no tumor lesions, were shown to the model with
patches being cropped randomly.
In addition to random cropping, random flipping on the coronal and sagittal plane was
applied as a augmentation technique during training.
A batch size of 6 and a learning rate of $10^{-3}$ were applied.
Each model was trained for 1000 epochs, with the number of warm up
epochs \cite{goyal2017accurate} set to 7.
Weights of the trained model of \cite{he2022masked},
developed on ImageNet, were used to initialize the transformer layers
in the encoder, while weights of the encoding layer and the decoder were
randomly initialized.
During test time a stride of size $64\times42\times2$ performing 6
repetitions was used to process whole \ac{mri} volumes.

\paragraph{Metrics}
We used voxel wise \ac{auroc} and \ac{ap} as performance measures.
Only voxels lying inside the breast tissue segmentation mask were
taken into account for computation, as injection of contrast media
leads also to an uptake in tissue lying outside the breast area,
ref. Figure \ref{fig:conv_sgmt} the Supplemental Materials.
However, for \ac{ap} one has to consider the large
imbalance between normal and abnormal tissue labels,
leading to an expected small baseline performance.
Moreover, ground truth annotations were only given in the
form of bounding boxes, depicting only a rough delineation of
tumor tissue with several ground truth \ac{tp} scores involved
that should in fact be \ac{tn}.
This has an higher impact on \ac{ap} than on \ac{auroc}, as
\ac{tp} scores are involved in precision and recall but not in
the \ac{fpr} of the ROC, which also takes \ac{tn} labels into account.
% -----------------------------------------------------------------------------------------
% Results 
% -----------------------------------------------------------------------------------------
\section{Results}
\begin{figure}[b!]
  \centering
  \includegraphics[width=0.9\textwidth]{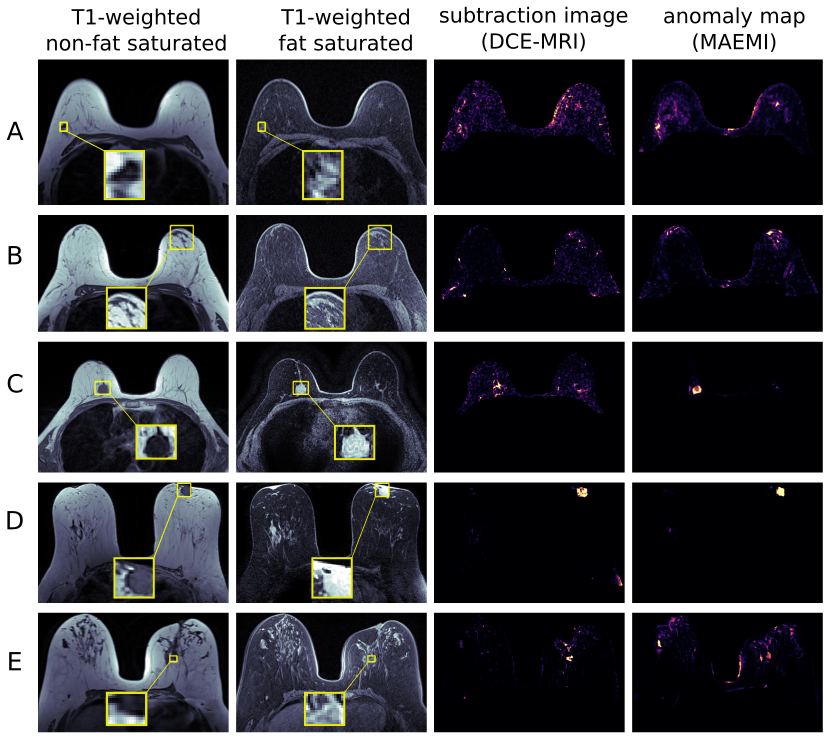}
  \caption{
  Example results. The first two columns show the non-contrast enhanced
  images used as an input to the anomaly detection model, and the
  last two columns present subtraction images generated by \ac{dce}-\ac{mri} and
  anomaly detection maps generated by \ac{maemi}, respectively.
  For patients in rows A, B and C, anomaly maps show superior performance over
  subtraction images.
  For patient D, both methods are able to identify the pathology.
  For patient E, our model only detects the borders of the pathology,
  while the subtraction image identifies the lesion.
  }
  \label{fig:examples}
\end{figure}

\Ac{vit}-patch size and masking ratio have been varied for hyperparameter tuning.
The best performing model featured a masking ratio of 90\% and a ViT-patch size of
$8\times8\times2$, \ac{auroc} and \ac{ap} results are shown in Table \ref{tab:results}.
Example results are shown in Figure \ref{fig:examples}.
\begin{table}[htb!]
    \begin{minipage}{.45\linewidth}
        \centering
        \caption{
            \Ac{maemi} achieves a higher \ac{auroc},
            while \ac{dce}-\ac{mri} features a higher
            \ac{ap}. However, significance of \ac{ap} results are limited
            (see Section \ref{sec:methods}).
        }
        \begin{tabular}{l|cc}
            & AUROC & AP \\
            \hline
            MAEMI & \textbf{0.732} & 0.081\\
            \ac{dce}-\ac{mri} & 0.705 & \textbf{0.127}\\ 
        \end{tabular}
        \label{tab:results}
    \end{minipage}%
    \hfill
    \begin{minipage}{.5\linewidth}
        \centering
        \caption{
            Ablation study on different (axial) ViT-patch sizes,
            for a fixed masking ratio of 90\%. Smaller patch sizes
            lead to higher performance.
        }
        \begin{tabular}{l|cc}
            ViT-patch size & AUROC & AP \\
            \hline
            $6\times6\times4$ & 0.724 & 0.0784\\
            $8\times8\times4$ & 0.712 & 0.0777\\
            $12\times12\times4$ & 0.660 & 0.0750\\
            $24\times24\times4$ & 0.546 & 0.0560
        \end{tabular}
        \label{tab:patch_size}
    \end{minipage} 
\end{table}
% images are taken from
% Breast_MRI_702
% Breast_MRI_353
% Breast_MRI_199
% Breast_MRI_402
% Breast_MRI_037
Mean baseline performance of the \ac{ap} measure, given by the number of
voxels inside the bounding box divided by the number of voxels lying
inside the breast tissue segmentation mask, was given by $0.046$.

\begin{figure}[htb!]
  \centering
  \includegraphics[width=0.7\textwidth]{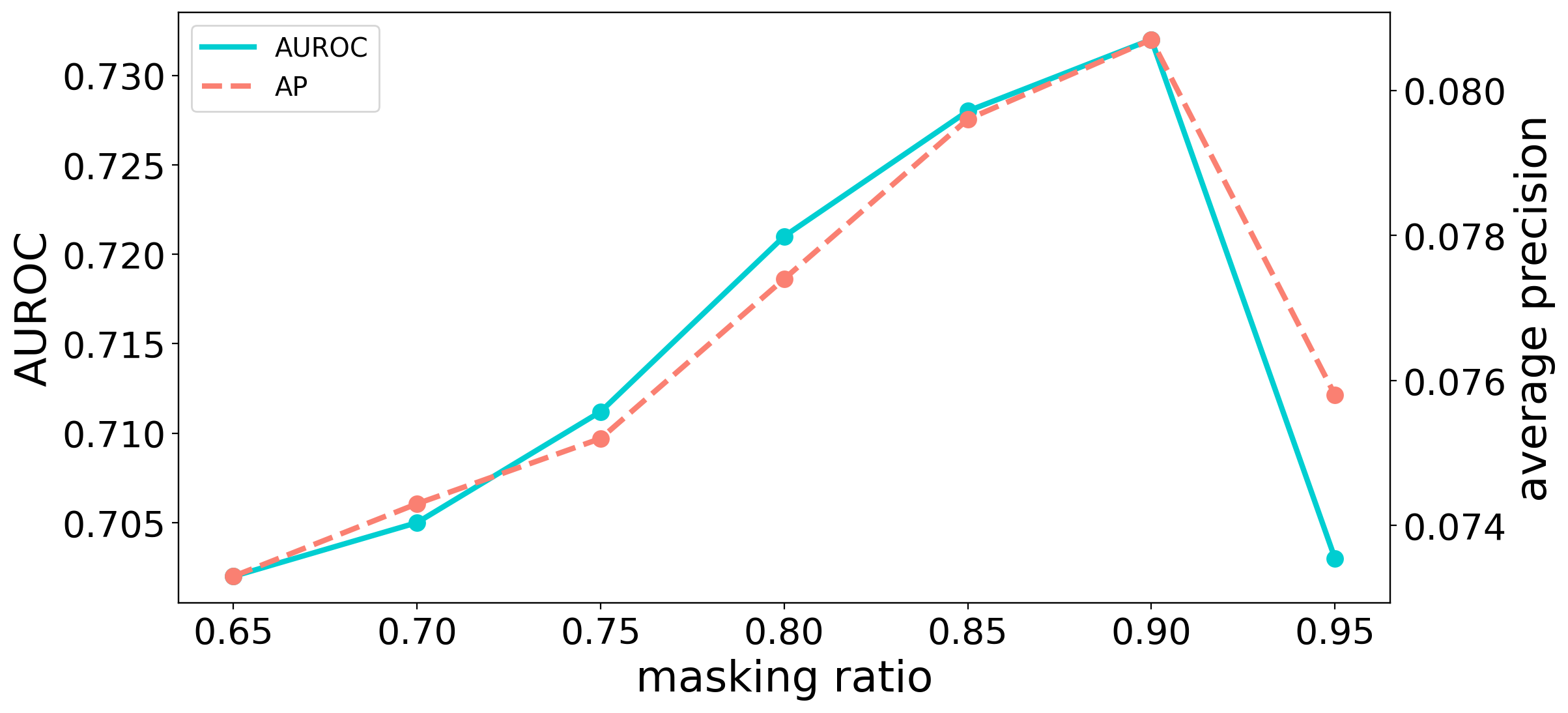}
  \caption{
  Ablation study on the masking ratio for a fixed ViT-patch
  size of $8 \times 8 \times 2$. High masking ratios lead to better performance,
  with an optimum reached at 90\%. Afterwards, performance suffers from
  a steep decline.
  }
  \label{fig:ablation}
\end{figure}

\paragraph{Ablation Studies}
We studied the influence of the patch size and masking ratio
on model performance.
Figure \ref{fig:ablation} presents the dependency of \ac{auroc} and \ac{ap}
on the masking ratio for a fixed ViT-patch size of $8\times8\times2$.
Dependency on ViT-patch size for a fixed masking ratio of 90\% is given in
Table \ref{tab:patch_size}.
The  Nvidia RTX A6000 used for training, featuring 48 GB of memory,
only allowed for a smallest size of $8\times8\times2$.
Therefore, the slice dimension of ViT-patches was fixed at a value of $4$ pixels,
probing only different axial sizes.

% -----------------------------------------------------------------------------------------
% Discussion 
% -----------------------------------------------------------------------------------------
\section{Discussion}
\label{sec:discussion}
We developed a new transformer-based \ac{ae} model that can be trained on multi-spectral volumetric
medical imaging data.
We have applied our model to anomaly detection on non-contrast
enhanced breast \ac{mri}.
Model performance was at the same level as for \ac{dce}-\ac{mri}
generated subtraction images.
Thus, we were able to  demonstrate the general ability for
automated identification of suspicious pathologies on non-contrast
enhanced breast \ac{mri}.

\Ac{maemi} achieved a higher \ac{auroc} while \ac{dce}-\ac{mri} generated
subtraction images resulting in better \ac{ap} performance.
Limitations of evaluation metrics due to ground truth labels given in the
form of bounding boxes were stated in Section \ref{sec:methods}.
Determination of an optimal performance measure for ground truth
bounding boxes in the case of unsupervised anomaly detection remains
an active area of research.

We found an optimal masking ratio of 90\% for our model.
\textit{Feichtenhofer et al.} \cite{feichtenhofer2022masked} identified
the same ratio to work best for video classification.
However, masking plays a significantly different role for \ac{mae}
architectures utilized for anomaly detection, with ViT-patch removal
not only applied during training but also during test time.
\textit{Tian et al.} \cite{tian2022unsupervised} employed the standard
masking ratio of 75\% for their model, trained to identify pathologies on
2D X-ray and colonoscopy data and did not report any
ablation studies.

Automated generation of anomaly maps from non-contrast enhanced imaging
allows for identification of suspicious lesions without the need for temporal imaging,
which results in a drastic reduction of costs and acquisition time,
two major factors limiting widespread application of \ac{mri} in breast
cancer screening \cite{leithner2019abbreviated}.
Therefore, we paved the way for a more widespread application of \ac{mri}
in breast cancer diagnosis.
Furthermore, patients can potentially be spared from intravenous injection of contrast media,
which is known to be able to cause side effects \cite{hasebroock2009toxicity}.
Clinical differences and benefits between \ac{maemi}
and \ac{dce}-\ac{mri} will still need to be investigated in a larger clinical study.

\subsubsection{Data Use Declaration}
All data used for this study is publicly available from
The Cancer Imaging Archive \cite{saha2021dynamic,clark2013cancer} under
the \href{https://creativecommons.org/licenses/by-nc/4.0/}{CC BY-NC 4.0} license.

% -----------------------------------------------------------------------------------------
% Appendix
% -----------------------------------------------------------------------------------------
\pagenumbering{gobble}
\appendix
\section*{Supplementary Material}
\label{sec:appendix}

\begin{figure}
  \centering
  \includegraphics[width=0.99\textwidth]{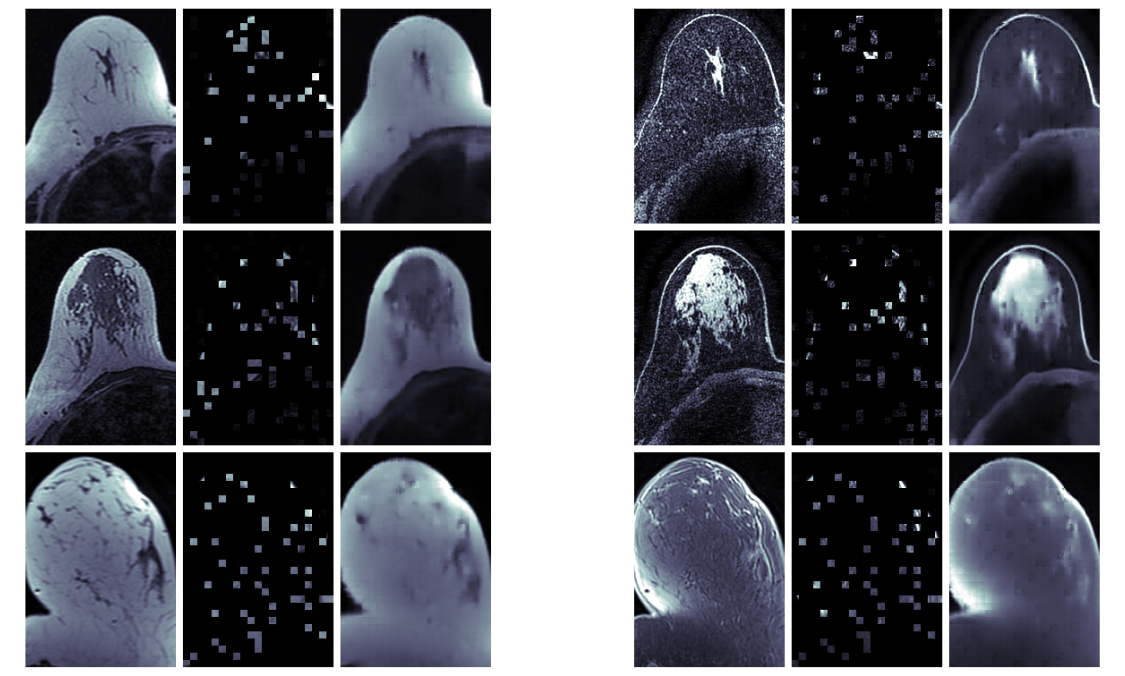}
  \caption{
  Reconstruction examples. The left block shows axial slices
  of T1 non-fat saturated MRI-patches
  and the right block T1 fat saturated slices.
  The first column shows unaltered MRI-patches, the
  second column the masked model input and the third column
  the MRI-patches recovered by \ac{maemi}.
  Examples represent a masking ratio of 90\%
  (for the whole 3D patch)
  and a ViT-patch size of $8\times8\times2$.
  }
  \label{fig:mask_examples}
\end{figure}
\begin{figure}
  \centering
  \includegraphics[width=0.8\textwidth]{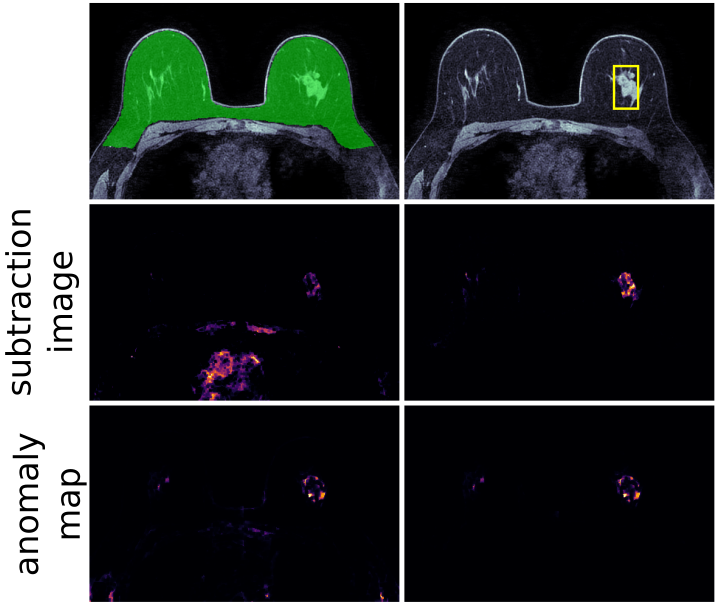}
  \caption{
    Subtraction images and anomaly maps were multiplied with
    segmentation masks to remove anomalies lying obviously outside
    of the breast tissue.
    This is mainly needed as contrast agent is also taken
    up in organs outside the breast.
    The left column shows the raw subtraction/anomaly map,
    and the right column the raw maps multiplied with the
    segmentation mask of the image in the upper left corner.
    Performance metrics were only calculated for voxels lying
    inside the segmentation mask, limiting the influence of
    trivial predictions, i.e. voxels that represent air do not
    containing any anomalies.
  }
  \label{fig:conv_sgmt}
\end{figure}
% images were taken from: Breast_MRI_581

\newpage

\bibliographystyle{splncs04}
\bibliography{references}

\begin{thebibliography}{10}
\providecommand{\url}[1]{\texttt{#1}}
\providecommand{\urlprefix}{URL }
\providecommand{\doi}[1]{https://doi.org/#1}

\bibitem{baur2021autoencoders}
Baur, C., Denner, S., Wiestler, B., Navab, N., Albarqouni, S.: {Autoencoders
  for unsupervised anomaly segmentation in brain MR images: a comparative
  study}. Medical Image Analysis  \textbf{69},  101952 (2021)

\bibitem{bercea2022federated}
Bercea, C.I., Wiestler, B., Rueckert, D., Albarqouni, S.: Federated
  disentangled representation learning for unsupervised brain anomaly
  detection. Nature Machine Intelligence  \textbf{4}(8),  685--695 (2022)

\bibitem{bergmann2018improving}
Bergmann, P., L{\"o}we, S., Fauser, M., Sattlegger, D., Steger, C.: Improving
  unsupervised defect segmentation by applying structural similarity to
  autoencoders. arXiv preprint arXiv:1807.02011  (2018)

\bibitem{ching2018opportunities}
Ching, T., Himmelstein, D.S., Beaulieu-Jones, B.K., Kalinin, A.A., Do, B.T.,
  Way, G.P., Ferrero, E., Agapow, P.M., Zietz, M., Hoffman, M.M., et~al.:
  {Opportunities and obstacles for deep learning in biology and medicine}.
  Journal of The Royal Society Interface  \textbf{15}(141),  20170387 (2018)

\bibitem{chow2020anomaly}
Chow, J.K., Su, Z., Wu, J., Tan, P.S., Mao, X., Wang, Y.H.: Anomaly detection
  of defects on concrete structures with the convolutional autoencoder.
  Advanced Engineering Informatics  \textbf{45},  101105 (2020)

\bibitem{UNET}
Chris, L.: {3D-Breast-FGT-and-Blood-Vessel-Segmentation}.
  \url{https://github.com/mazurowski-lab/3D-Breast-FGT-and-Blood-Vessel-Segmentation}
  (2022)

\bibitem{clark2013cancer}
Clark, K., Vendt, B., Smith, K., Freymann, J., Kirby, J., Koppel, P., Moore,
  S., Phillips, S., Maffitt, D., Pringle, M., et~al.: {The Cancer Imaging
  Archive (TCIA): maintaining and operating a public information repository}.
  Journal of digital imaging  \textbf{26},  1045--1057 (2013)

\bibitem{dosovitskiy2020image}
Dosovitskiy, A., Beyer, L., Kolesnikov, A., Weissenborn, D., Zhai, X.,
  Unterthiner, T., Dehghani, M., Minderer, M., Heigold, G., Gelly, S., et~al.:
  {An image is worth 16x16 words: Transformers for image recognition at scale}.
  arXiv preprint arXiv:2010.11929  (2020)

\bibitem{feichtenhofer2022masked}
Feichtenhofer, C., Fan, H., Li, Y., He, K.: {Masked autoencoders as
  spatiotemporal learners}. arXiv preprint arXiv:2205.09113  (2022)

\bibitem{gong2019memorizing}
Gong, D., Liu, L., Le, V., Saha, B., Mansour, M.R., Venkatesh, S., Hengel,
  A.v.d.: {Memorizing normality to detect anomaly: Memory-augmented deep
  autoencoder for unsupervised anomaly detection}. In: Proceedings of the
  IEEE/CVF International Conference on Computer Vision. pp. 1705--1714 (2019)

\bibitem{goyal2017accurate}
Goyal, P., Doll{\'a}r, P., Girshick, R., Noordhuis, P., Wesolowski, L., Kyrola,
  A., Tulloch, A., Jia, Y., He, K.: {Accurate, large minibatch sgd: Training
  imagenet in 1 hour}. arXiv preprint arXiv:1706.02677  (2017)

\bibitem{hasebroock2009toxicity}
Hasebroock, K.M., Serkova, N.J.: {Toxicity of MRI and CT contrast agents}.
  Expert opinion on drug metabolism \& toxicology  \textbf{5}(4),  403--416
  (2009)

\bibitem{he2022masked}
He, K., Chen, X., Xie, S., Li, Y., Doll{\'a}r, P., Girshick, R.: {Masked
  autoencoders are scalable vision learners}. In: Proceedings of the IEEE/CVF
  Conference on Computer Vision and Pattern Recognition. pp. 16000--16009
  (2022)

\bibitem{herent2019detection}
Herent, P., Schmauch, B., Jehanno, P., Dehaene, O., Saillard, C., Balleyguier,
  C., Arfi-Rouche, J., J{\'e}gou, S.: {Detection and characterization of MRI
  breast lesions using deep learning}. Diagnostic and interventional imaging
  \textbf{100}(4),  219--225 (2019)

\bibitem{kascenas2022denoising}
Kascenas, A., Pugeault, N., O'Neil, A.Q.: {Denoising autoencoders for
  unsupervised anomaly detection in brain MRI}. In: International Conference on
  Medical Imaging with Deep Learning. pp. 653--664. PMLR (2022)

\bibitem{leithner2019abbreviated}
Leithner, D., Moy, L., Morris, E.A., Marino, M.A., Helbich, T.H., Pinker, K.:
  {Abbreviated MRI of the breast: does it provide value?} Journal of Magnetic
  Resonance Imaging  \textbf{49}(7),  e85--e100 (2019)

\bibitem{maicas2017deep}
Maicas, G., Carneiro, G., Bradley, A.P., Nascimento, J.C., Reid, I.: {Deep
  reinforcement learning for active breast lesion detection from DCE-MRI}. In:
  Medical Image Computing and Computer Assisted Intervention- MICCAI 2017: 20th
  International Conference, Quebec City, QC, Canada, September 11-13, 2017,
  Proceedings, Part III. pp. 665--673. Springer (2017)

\bibitem{prabhakar2023vit}
Prabhakar, C., Li, H.B., Yang, J., Shit, S., Wiestler, B., Menze, B.:
  {ViT-AE++: Improving Vision Transformer Autoencoder for Self-supervised
  Medical Image Representations}. arXiv preprint arXiv:2301.07382  (2023)

\bibitem{saha2021dynamic}
Saha, A., Harowicz, M., Grimm, L., Weng, J., Cain, E., Kim, C., Ghate, S.,
  Walsh, R., Mazurowski, M.: {Dynamic contrast-enhanced magnetic resonance
  images of breast cancer patients with tumor locations}. The Cancer Imaging
  Archive  (2021)

\bibitem{saha2018machine}
Saha, A., Harowicz, M.R., Grimm, L.J., Kim, C.E., Ghate, S.V., Walsh, R.,
  Mazurowski, M.A.: {A machine learning approach to radiogenomics of breast
  cancer: a study of 922 subjects and 529 DCE-MRI features}. British journal of
  cancer  \textbf{119}(4),  508--516 (2018)

\bibitem{schwartz2022maeday}
Schwartz, E., Arbelle, A., Karlinsky, L., Harary, S., Scheidegger, F., Doveh,
  S., Giryes, R.: {MAEDAY: MAE for few and zero shot AnomalY-Detection}. arXiv
  preprint arXiv:2211.14307  (2022)

\bibitem{somepalli2021unsupervised}
Somepalli, G., Wu, Y., Balaji, Y., Vinzamuri, B., Feizi, S.: {Unsupervised
  anomaly detection with adversarial mirrored autoencoders}. In: Uncertainty in
  Artificial Intelligence. pp. 365--375. PMLR (2021)

\bibitem{sun2017revisiting}
Sun, C., Shrivastava, A., Singh, S., Gupta, A.: Revisiting unreasonable
  effectiveness of data in deep learning era. In: Proceedings of the IEEE
  international conference on computer vision. pp. 843--852 (2017)

\bibitem{tian2022unsupervised}
Tian, Y., Pang, G., Liu, Y., Wang, C., Chen, Y., Liu, F., Singh, R., Verjans,
  J.W., Carneiro, G.: {Unsupervised anomaly detection in medical images with a
  memory-augmented multi-level cross-attentional masked autoencoder}. arXiv
  preprint arXiv:2203.11725  (2022)

\bibitem{tschuchnig2022anomaly}
Tschuchnig, M.E., Gadermayr, M.: Anomaly detection in medical imaging-a mini
  review. In: Data Science--Analytics and Applications: Proceedings of the 4th
  International Data Science Conference--iDSC2021. pp. 33--38. Springer (2022)

\bibitem{turnbull2009dynamic}
Turnbull, L.W.: {Dynamic contrast-enhanced MRI in the diagnosis and management
  of breast cancer}. NMR in Biomedicine: An International Journal Devoted to
  the Development and Application of Magnetic Resonance In Vivo
  \textbf{22}(1),  28--39 (2009)

\bibitem{Wyatt_2022_CVPR}
Wyatt, J., Leach, A., Schmon, S.M., Willcocks, C.G.: {AnoDDPM: Anomaly
  Detection With Denoising Diffusion Probabilistic Models Using Simplex Noise}.
  In: Proceedings of the IEEE/CVF Conference on Computer Vision and Pattern
  Recognition (CVPR) Workshops. pp. 650--656 (June 2022)

\bibitem{yan2021learning}
Yan, X., Zhang, H., Xu, X., Hu, X., Heng, P.A.: Learning semantic context from
  normal samples for unsupervised anomaly detection. In: Proceedings of the
  AAAI Conference on Artificial Intelligence. vol.~35, pp. 3110--3118 (2021)

\bibitem{zavrtanik2021reconstruction}
Zavrtanik, V., Kristan, M., Sko{\v{c}}aj, D.: Reconstruction by inpainting for
  visual anomaly detection. Pattern Recognition  \textbf{112},  107706 (2021)

\end{thebibliography}

\end{document}